\documentclass[]{interact}
\usepackage{epstopdf}% To incorporate .eps illustrations using PDFLaTeX, etc.
\usepackage[caption=false]{subfig}% Support for small, `sub' figures and tables

\usepackage[numbers,sort&compress]{natbib}
\usepackage{tikz}
\usepackage{geometry}
\usepackage{hyperref}
\hypersetup{colorlinks=true, citecolor=blue} 
\usepackage[ruled,linesnumbered]{algorithm2e}

\newcommand{\balpha}{\pmb{\alpha}}
\newcommand{\btheta}{\pmb{\theta}}
\newcommand{\bbeta}{\pmb{\beta}}
\newcommand{\bow}{\pmb{w}}
\newcommand{\ba}{\pmb{a}}
\newcommand{\bdelta}{\pmb{\delta}}

\newcommand{\bgamma}{\pmb{\gamma}}

\newcommand{\bh}{\pmb{h}}

\newcommand{\bkappa}{\pmb{\kappa}}

\newcommand{\bt}{\pmb{t}}

\newcommand{\by}{\pmb{y}}

\newcommand{\bw}{\pmb{w}}
\newcommand{\bW}{\pmb{W}}

\DeclareMathOperator{\DT}{DT}
\DeclareMathOperator{\DN}{DN}
\newcommand{\ORD}{\mathrm{ORD}}
\newcommand{\SCH}{\mathrm{SCH}}
\newcommand{\PWE}{\mathrm{PWE}}
\newcommand{\HOL}{\mathrm{HOL}}

\newcommand{\test}{\mathrm{test}}
\newcommand{\train}{\mathrm{train}}
\newcommand{\PE}{\mathrm{PE}}

\usetikzlibrary{shapes,arrows}
\usetikzlibrary{shapes.geometric}
\tikzset{database/.style={cylinder,aspect=0.5,draw,rotate=90,path picture={
\draw (path picture bounding box.160) to[out=180,in=180] (path picture bounding
box.20);
\draw (path picture bounding box.200) to[out=180,in=180] (path picture bounding
box.340);
}}}

\definecolor{orange}{RGB}{253,192,134}
\definecolor{violet}{RGB}{190,174,212}
\definecolor{green}{RGB}{127,201,127}

\bibpunct[, ]{[}{]}{,}{n}{,}{,}% Citation support using natbib.sty
\renewcommand\bibfont{\fontsize{10}{12}\selectfont}% Bibliography support using natbib.sty

\theoremstyle{plain}% Theorem-like structures provided by amsthm.sty

\theoremstyle{definition}

\theoremstyle{remark}

\begin{document}

\title{Bayesian hierarchical models for the prediction of the driver flow and passenger waiting times in a stochastic carpooling service}

\author{
\name{Panayotis Papoutsis \textsuperscript{a,b,c}\thanks{CONTACT Panayotis Papoutsis. Email: papoutsispanayotis@gmail.com}, Bertrand Michel\textsuperscript{a,b}, Anne Philippe \textsuperscript{b} and Tarn Duong\textsuperscript{c}}
\affil{\textsuperscript{a}Department of Computing and Mathematics, Nantes Central Engineering School, F-44300, France; \textsuperscript{b}Jean Leray Mathematics Laboratory, University of Nantes, F-44300, France;\textsuperscript{c}Ecov, F-44200, France}}

\maketitle

\begin{abstract}
Carpooling is an integral component in smart carbon-neutral cities, in particular to facilitate home-work commuting. We study an innovative carpooling service developed by the start-up Ecov which specialises in home-work commutes in peri-urban and rural regions. When a passenger makes a carpooling request, a designated driver is not assigned as in a traditional carpooling service; rather the passenger waits for the first driver, from a population of non-professional drivers who are already en route, to arrive. We propose a two-stage Bayesian hierarchical model to overcome the considerable difficulties, due to the sparsely observed driver and passenger data from an embryonic stochastic carpooling service, to deliver high-quality predictions of driver flow and passenger waiting times. The first stage focuses on the driver flows, whose predictions are aggregated at the daily level to compensate the data sparsity. The second stage processes this single daily driver flow into sub-daily (e.g. hourly) predictions of the passenger waiting times. We demonstrate that our model mostly outperforms frequentist and non-hierarchical Bayesian methods for observed data from operational carpooling service in Lyon, France and we also validated our model on simulated data.  

\end{abstract}


\section{Introduction}

Providing ecologically sustainable transportation that is accessible for all is one of the key challenges in the transition to post-carbon societies. An innovative solution devised by the French start-up, Ecov (\url{wwww.ecov.fr}), is a carpooling service for rural and urban peripheral regions. These regions are often neglected by the start-up sector as it tends to focus on technology-savvy populations in dense urban regions. These carpooling services, whilst having been initiated in the private sector, are developed and operated in close collaboration with local government authorities in order to satisfy the mobility requirements in these marginalised areas with sparser population and physical/digital infrastructure. The key innovation brought to the market by Ecov is the provision of carpooling lines, which closely resemble traditional bus lines. These carpooling lines link physical meeting points between which carpooling is assured at suitable regularity. This concentrates the demand and the supply of carpooling to reach a critical mass more quickly and more sustainable. The meeting points are placed strategically in highly frequented areas, which take into account various factors such as aggregated traffic flow, socioeconomic characteristics, pedestrian accessibility, local government regulations, etc. Pick ups and drop-offs at other locations than these meeting points are not facilitated by the provider, though they are not disallowed, which ensure more flexibility than bus lines. The meeting points resemble bus shelters, except where a passenger waiting for a bus usually only requires a simple hand gesture to the driver to indicate that they intend to embark, the carpooling passenger must make an explicit carpooling request on an electronic console. This request is then displayed on a electronic sign on the roadside which informs \textit{all} passing drivers of a passenger request to a specified destination. This driver is not allocated in advance -- this real-time, stochastic matching between a passenger and driver, is a major distinguishing feature of Ecov carpooling services in contrast to their competitors (such as Uber, Lyft, Kapten etc). It is this stochastic matching between a passenger and a flow of potential drivers, along with the aggregating effects of the physical meeting points, that enable carpooling to reach economical feasibility in sparsely populated regions. 

The stochastic matching from a mathematical and technological point-of-view is more difficult than the deterministic passenger-driver matching in order to provide a reliable waiting time of a driver arrival. In the latter, a reliable waiting time for a passenger request requires only the tracking of an allocated driver, whereas stochastic matching requires both (a) the tracking of multiple potential drivers and (b) an understanding of the general driver flow. The digital technological infrastructure is key in delivering reliable waiting times to passengers. Ecov provides users with a mobile phone application for their carpooling services: passengers receive updates about the waiting time for a driver arrival, and drivers receive notifications of passengers waiting at the meeting points, and crucially, are able to share their GPS locations in real-time with Ecov. These driver GPS traces, by providing pertinent information, ensure the quality of the carpooling service. This information includes the daily driver flow and the passenger waiting time, which we focus on in this paper.

Due the complexity of the relationship between the driver GPS traces and the passenger waiting times, and the scarcity of the observed data due to the novelty of the stochastic carpooling, we propose a hierarchical approach where we first build predictive models of the potential driver flow from the observed GPS driver traces. At the time when a passenger request is made, we do not have a sufficiently detailed knowledge of the instantaneous potential driver flow, so we model this driver flow first as a moving average of previous driver flows. Then we model the passenger waiting time as a regression model with covariates based on the driver traffic flow modelled in the first stage.

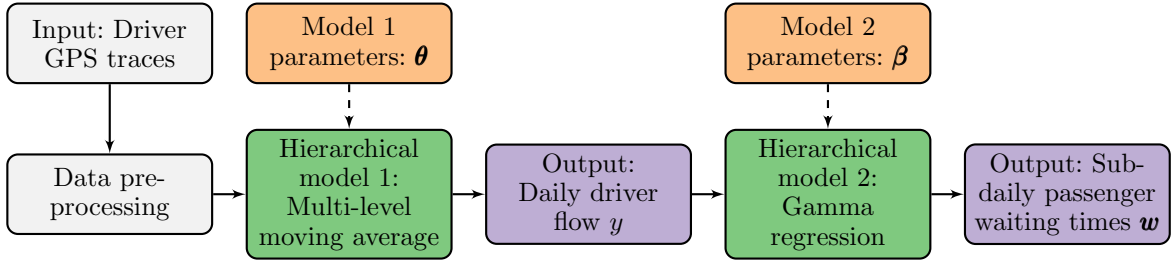
\begin{figure}[!ht]
\centering
% Define block styles
\tikzstyle{block} = [rectangle, 
        draw, 
        thick,
        text width=7em, 
        text centered, 
        rounded corners,
        minimum height=3em,
        node distance=2cm]

\tikzstyle{line} = [draw, -latex']

\begin{small}
\begin{tikzpicture}[node distance = 3cm, auto]
    % Place nodes
   \node [block,
        minimum width=2cm,
        fill=gray!10] 
        (input) {Input: Driver \\ GPS traces};
        
    \node [block,
        fill=gray!10,
        below of=input] 
        (preprocessing) {Data preprocessing};
    
    \node [block,
        fill=green, 
        right of=preprocessing,
        node distance=3.15cm] 
        (multilevel) {Hierarchical model~1: Multi-level moving average};
    
    \node [block,
        fill=violet,
        right of=multilevel,
        node distance=3.15cm]
        (first) {Output: \\ Daily driver flow $y$};
        
    \node [block,
        fill=green,
        right of=first,
        node distance=3.15cm] 
        (gamma) {Hierarchical model~2: \\ Gamma \\ regression};
        
    \node [block,
        fill=violet,
        right of=gamma,
        node distance=3.15cm] 
        (second) {Output: Sub-daily passenger waiting times $\bw$};
        
    \node[block,
        fill=orange,
        above of=multilevel]
        (theta) {Model 1 \\ parameters: $\btheta$}; 
        
    \node[block,
        fill=orange,
        above of=gamma]
        (beta) {Model 2 \\ parameters: $\bbeta$};
    
    % Draw edges
    \path [line,thick] (input) -- (preprocessing);
    \path [line,thick] (preprocessing) -- (multilevel);
    \path [line,thick] (multilevel) -- (first);
    \path [line,thick] (first) -- (gamma);
    \path [line,thick] (gamma) -- (second);
    \path [line,dashed,thick] (theta) -- (multilevel);
    \path [line,dashed,thick] (beta) -- (gamma);
\end{tikzpicture}
\end{small}

\caption{Flowchart of Bayesian hierarchical model for driver flow and passenger waiting time prediction. The input data (driver GPS traces) are in grey, the hierarchical models in green, the model parameters in orange, and the model outputs in purple.}
\label{fig:model_structure}
\end{figure}

In the flowchart in Figure~\ref{fig:model_structure}, our Bayesian multi-level hierarchical model is composed of two nested stages. The input data (driver GPS traces) are preprocessed, as outlined in \cite{papoutsis2020door}, so that they are suitable as subsequent input into the hierarchical models themselves. The first model is a multi-level moving average model whose coefficients $\btheta$ with levels depending on if the current type of day: working, weekend, public or school holiday. Bayesian multi-level models crucially are able to suitably model the driver flows with these overlapping levels (e.g. the driver flow for public holiday which is also a school holiday is different to that of public holidays outside of the school holiday period).  These levels are known to be highly influential \citep{bao2017investigation}.The output from the first hierarchical model is the daily driver flow, which is the immediate input to the second hierarchical model. The latter is a Gamma regression, whose regression coefficient $\bbeta$ has $S$ components for each of the time intervals into which a 24-hour period is divided. The role of $\bbeta$ is to assign the daily traffic flow to each of these sub-daily time intervals.  The output of this second hierarchical model is the temporal profile of the passenger waiting times $\bw$ for each of the sub-daily time intervals. The scarcity of the input data (driver GPS traces) only allows us to model the driver flow robustly at a daily level, whereas a higher temporal resolution of the output passenger waiting times is required for a carpooling service. Bayesian hierarchical models offer an intuitive treatment of these differing temporal resolutions within a single workflow.

In Section~\ref{sec:traffic_flow_model} the first stage of the hierarchical model for the daily driver traffic flow is described. In Section~\ref{sec:waiting_times_model} the second stage of the hierarchical model for the passenger waiting times is described. In Section~\ref{sec:validation} we validate our model on simulated data and then compare itself performance with frequentist and non-hierarchical Bayesian models on empirical data from an operational carpooling service. We end the paper with a discussion and some future perspectives.

\section{Bayesian multi-level moving averages of the daily driver flow} \label{sec:traffic_flow_model}

\begin{figure}[!ht]
\centering
\includegraphics[height=0.3\textwidth]{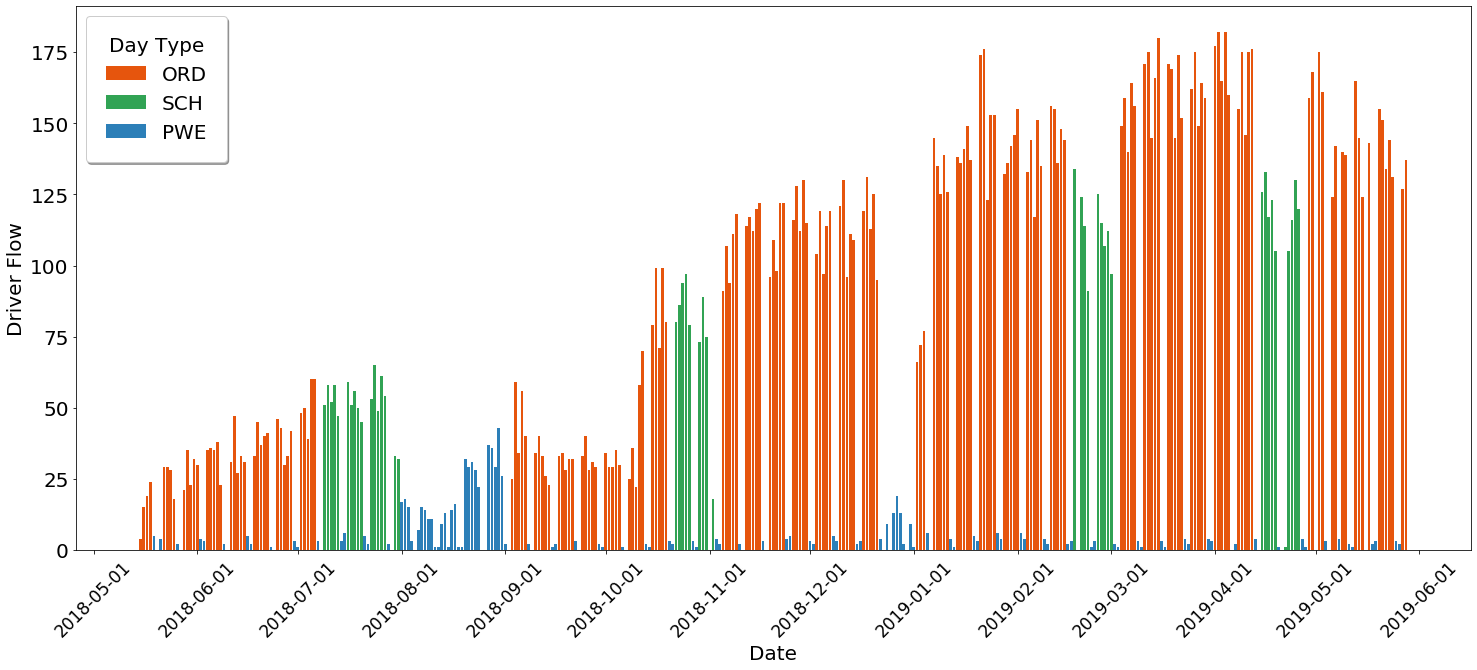} 
\includegraphics[height=0.3\textwidth]{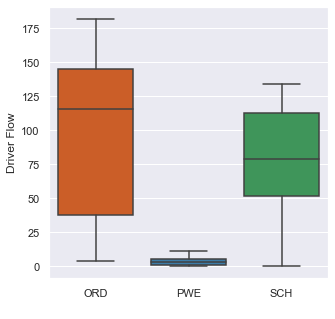} 
\caption{Daily driver flow on the Lane carpooling network, from 2018-05-15 to 2019-05-31. Left. Daily times series. Right. Aggregate driver flow by day type. The ordinary work days (ORD) are in orange, the school holidays (SCH) in green and the public holidays/weekend days (PWE) in blue.}
\label{fig:vehicule_counts_daytypes}
\end{figure}

As the driver flow is a fundamental quantity in transportation research, its estimation/prediction is the subject of a vast field of active research so we cite only those references with a direct connection with the analysis presented in this paper. Historically the simplest models are the moving window averages, see for instance \cite{edes1980improved}.  More advanced methods draw from time series analysis, within a frequentist \citep{ding2002traffic} or a Bayesian framework \citep{ghosh2007bayesian} %, as well as from machine learning within a deep learning framework \citep{gu2019improved}, 
have been posited. Our proposed approach of a Bayesian multi-level moving average is a combination of the approaches of \cite{edes1980improved} and \cite{ghosh2007bayesian} which combines the robustness and simplicity of moving averages, with the targeted adjustments of multi-level coefficients.
The empirical data in this paper are extracted from the Lane stochastic carpooling service (\url{lanemove.com}) operated by Ecov, in conjunction with Instant System (\url{instant-system.com}), since May 2018 in the south-eastern peri-urban regions around Lyon, France. See \cite{papoutsis2020door} for more details on its set-up. We focus on the driver GPS traces for the 382 days from 2018-05-15 (service launch) to 2019-05-31 (beginning of the following year's summer holiday season in France). The daily driver flows in the Lane network are presented in Figure~\ref{fig:vehicule_counts_daytypes}, where we enumerate each trajectory, rather than each unique driver. So a single driver can make several trajectories within this time period.  The colour coding is induced by the day type, defined as 
\begin{align}
\label{eq:day_types_mapping}
\DT(i) = \begin{cases}
\ORD  & \mathrm{if \ day} \ i \ \mathrm{is \ an \ ordinary \ work day} \\
\SCH  & \mathrm{if \ day} \ i \ \mathrm{is \ a \ school \ holiday}\\
\PWE  & \mathrm{if \ day} \ i \ \mathrm{is \ a \ public \ holiday \ or \ a \ weekend}
\end{cases}
\end{align}  
where $i$ is the index of the day from the service launch, i.e. $i=1$ for 2018-05-15 etc. In  Figure~\ref{fig:vehicule_counts_daytypes}, the work days are in orange, the school holidays in green, and the public holidays/weekends in blue. The classic temporal cycles of driver flow data in the left panel are present which indicate that a moving average is relevant approach for prediction. According to \cite{kung2014exploring,bao2017investigation} these day types are a key determinant of home-work daily commutes, which is verified by the box plots of the daily driver flow by day type in the right panel. The daily driver flow for a work day approaches 200 trajectories, which is about four times larger than driver flow on school holidays, and more than 10 times larger than on the public holiday/weekends.

\subsection{Multi-level moving average model specification}
From visual inspection of the daily driver flow in Figure~\ref{fig:vehicule_counts_daytypes} , a standard moving average which ignores these day types would be unable to account for the abrupt differences in driver flow when consecutive days are of different day types.  The recurrence relation of the daily driver flow $y_i$, on day $i \geq K \geq 1$, satisfies 
\begin{eqnarray}
\label{eq:traffic_model}
    y_i & = & \alpha_{\DT(i)} \sum_{k=1}^K  \eta_{\DT(i-k)} y_{i-k} + \varepsilon_i \\
        & = & \alpha_{\DT(i)} \sum_{k=1}^K \left[ \mathbf{1}\{\DT(i-k)  = \ORD\} + \eta_{\SCH} \mathbf{1}\{\DT(i-k)  = \SCH\} \right. \notag \\
        & &  + \left. \eta_{\PWE} \mathbf{1} \{\DT(i-k)  = \PWE\} \right] y_{i-k} + \varepsilon_i  \notag
\end{eqnarray}
where $\alpha_{\DT(\cdot)}$ is the coefficient for the current day $i$ and $\eta_{\DT(\cdot)}$ are the coefficients for the past $K$ driver flows and $\varepsilon_i, i=1, 2, \dots$ are a sequence of independent normal random variables $\mathcal{N}(0, \sigma_\varepsilon^2)$. To ensure identifiability of $\eta_{\DT(\cdot)}$, without loss of generality, we set $\eta_{\ORD} = 1$ for all days. To reduce the mathematical complexity, we assume that the ratios of the mean daily driver flow of the three day types to each other ($\bar{y}_{i,\ORD} : \bar{y}_{i,\SCH} : \bar{y}_{i,\PWE} $) remain constant for all days $i$ in the entire time period. 
The model in Equation~\eqref{eq:traffic_model} has a moving average structure of order $K$, but with two additional multi-level coefficients that make the average adaptive to the day types for the current day $i$ and the previous $K$ days. For example, if day $i$ is a school holiday, then the right hand side of Equation~\eqref{eq:traffic_model} is 
$\alpha_{\SCH} \sum_{k=1}^K [ \mathbf{1}\{\DT(i-k) = \ORD\} + \eta_{\SCH} \mathbf{1}\{\DT(i-k) = \SCH\} +
\eta_{\PWE} \mathbf{1}\{\DT(i-k)  = \PWE\} ] y_{i-k}$. In the summand, the day type indicator functions allows us to sum over the $K$ previous days, even if they are of different types. If a previous day is a work day, then its contribution to the current driver flow is $\alpha_{\SCH} y_{i-k}$; if a previous day is a school holiday then it is $\alpha_{\SCH} \eta_{\SCH} y_{i-k}$; if a previous day is a public holiday/weekend then it is $\alpha_{\SCH} \eta_{\PWE} y_{i-k}$. 
The first multi-level coefficient $\alpha_{\SCH}$ models the current driver flow, conditionally on its day type. The second set of multi-level coefficients $\eta_{\SCH}, \eta_{\PWE}$ re-scale the previous driver flows, assuming that the ratio of the flows of different day types is constant for all days. 

Our model in Equation~\eqref{eq:traffic_model} possesses a similar structure to an autoregressive model, though it does not strictly satisfy the definition of one. It cannot be defined with a back shift operator due to the action of the multi-level coefficients $\alpha_{\DT(\cdot)}$ and $\eta_{\DT(\cdot)}$, and the process $\{y_i, i=1, 2, \dots\}$ is non-stationary due to the drift in the driver participation rate after the launch of the carpooling service.

The multi-level model in Equation~\eqref{eq:traffic_model} is equally valid for frequentist or Bayesian approaches for parameter estimation. We adopt a Bayesian approach, in line with \cite{gelman2006data}. Let $\btheta = (\alpha_{\ORD}, \alpha_{\SCH}, \alpha_{\PWE}, \eta_{\ORD}, \eta_{\SCH}, \eta_{\PWE},\sigma_\varepsilon^{2})$ though recall that we fix $\eta_{\ORD} = 1$ identically. Suppose that we have $N$ days of observed daily driver flows $y_i, i=1, \dots, N$, where $N > K$, the order of the moving average. Since the error variables are independent Gaussian, then the conditional likelihood of $\by = (y_K, y_{K+1}, \dots, y_N)$ is  
\begin{align*}
L(\by|\btheta) =  \frac{1}{(2 \pi \sigma_\varepsilon^2)^{(N-K+1)/2}} \exp \left[ -\frac{1}{2\sigma_\varepsilon^2} \sum \limits_{i=K}^N (y_i-g_i(\btheta))^2 \right]
\end{align*}
where $g_i(\btheta) =\alpha_{\DT(i)} \sum_{k=1}^K  \eta_{\DT(i-k)} y_{i-k}$. This conditional likelihood is  formed by the product of the conditional densities of $y_i$ given $y_{i-K}, \dots, y_{i-1}$ for $i=K+1, \dots, N$.

In Bayesian analysis the parameter of interest $\btheta$ is a random variable, and its prior distribution $\pi$ represents our belief in its uncertainty. The posterior density $\pi({\btheta|\by})$ represents an update of the prior distribution by taking into account the observed data: 
$ \pi ({\btheta|\by}) \propto L(\by|\btheta)\pi(\btheta)$. 

In our case, we do not have access to existing knowledge that would provide an informative prior and thus we form a non-informative prior on $\btheta$, i.e.  $\pi(\btheta) \propto \sigma_\varepsilon^{-2}$ \citep[Chapter~1]{congdon2014applied}. This leads to the following posterior distribution
\begin{align}\label{eq:posterior_distribution}
    \pi({\btheta|\by}) & \propto  L(\by|\btheta)\pi(\btheta) \propto \frac{1}{\sigma_\varepsilon^{N-K+1}} \exp \left[ -\frac{1}{2\sigma_\varepsilon^2} \sum \limits_{i=K}^N (y_i-g_i(\btheta))^2 \right].
\end{align}
For the inference on $\btheta$, Monte Carlo approximations are require since the posterior distribution (and its moments, quantiles etc.) cannot be calculated explicitly. 
The most widely used family of methods is the Monte Carlo Markov Chain (MCMC) which aim to generate a Markov Chain $\{ \btheta_0,\btheta_1, \dots \}$ whose equilibrium distribution converges to the  posterior distribution $\pi({\btheta | \by})$. 

The next stage is to predict a driver flow $\tilde y$ in the future from the observed past data $\by$.  Bayesian prediction is based on the  posterior predictive distribution, that is the distribution of $\tilde y$ conditional on the observed past data $\by$. Its density $p(\tilde y|\by)$ 
is given by 
\begin{align}\label{eq:pred_distri_y}
    p( \tilde y|\by ) = \int_{\Theta} p( \tilde y|\btheta ,\by) \pi(\btheta|\by) \, d\btheta. 
\end{align}
Since $p(\tilde y|\by)$ is a compound probability distribution,  we can easily simulate samples from this predictive distribution. 

\subsection{Simulation algorithms for the daily driver flow}

We begin with defining the daily driver flow recurrence with no day types.  So Equation~\eqref{eq:traffic_model} with day types simplifies to 

\begin{align*}
    y_i & = \alpha \sum_{k=1}^K  y_{i-k} + \varepsilon_i.
\end{align*}
Since the multi-level coefficients for the day types are no longer present, this is indeed an autoregressive model. 
Algorithm~\ref{alg:traffic_simple} simulates a driver flow for a single day with no day types. 
The inputs are the day $i$, the coefficient $\alpha$, the autoregression order $K$, and the error variance $\sigma^2_\varepsilon$. The output is a single driver flow for day $i$. The repeat loop ensures that the simulated driver flow is strictly positive. To simulate a sequence of $N$ driver flows, we initialise the values generated by Algorithm~1 for $i=1, \dots, K$ days, and then iterate Algorithm~1 sequentially for $i=K+1, \dots, N$.

\begin{algorithm}[!ht]
\textbf{procedure} \textsc{TrafficFlow}($i, \alpha,K, \sigma^2_\varepsilon$) \\
\eIf{i <=  K }{ 
    initialise $y \longleftarrow \mathcal{N}(30, \sigma^2_\varepsilon)$
   }{
   \Repeat{$y > 0$}{
   $y \longleftarrow \mathcal{N}(\alpha \sum_{k=1}^K \textsc{TrafficFlow}(i-k, \alpha,K,\sigma^2_\varepsilon$)$,\sigma^2_\varepsilon)$
  }
}
\textbf{return}: $y$ driver flow for day $i$
\caption{Daily driver flow without day types}\label{alg:traffic_simple}
\end{algorithm}

With Algorithm~\ref{alg:traffic_simple} defined, it is straightforward to define one with day types (i.e. Equation~\eqref{eq:traffic_model}) in Algorithm~\ref{alg:traffic_flow_daytypes}. This has similar inputs the day $i$, the day type coefficients $\btheta$, the autoregression order $K$, the error variance $\sigma^2_\varepsilon$, and except that the scalar $\alpha$ is replaced with the vector coefficients $\btheta$. The output is the daily driver flow for day $i$, accounting for the day types of the days preceding day $i$.  

\begin{algorithm}[!ht]
\textbf{procedure} \textsc{TrafficFlowDT}($i, \btheta, K, \sigma^2_\varepsilon$) \\
    \eIf{$\DT(i) == \ORD$}
    {$y \longleftarrow \textsc{TrafficFlow}(i, \alpha_{\ORD}, K, \sigma^2_\varepsilon)$}
    {
     \eIf{$\DT(i) == \SCH$}
     {$y \longleftarrow  \textsc{TrafficFlow}(i, \alpha_{\SCH} \eta_{\SCH}, K, \sigma^2_\varepsilon)$}
     {
       \If{$\DT(i) == \PWE$}
       {$y \longleftarrow \textsc{TrafficFlow}(i, \alpha_{\PWE} \eta_{\PWE}, K, \sigma^2_\varepsilon)$}
     }
    }
 \textbf{return}: $y$ driver flow for day $i$

\caption{Daily driver flow with day types}
\label{alg:traffic_flow_daytypes}
\end{algorithm}

For the choice of an MCMC sampler, we use the NUT sampler \citep{hoffman2014no}. The NUT sampler is used by default in the pyStan package (\url{https://pystan.readthedocs.io}), a Python interface to Stan (\url{https://mc-stan.org}), which is a state-of-art platform for Bayesian computations, amongst other functionalities.To carry out the complicated integration and then a random draw from the posterior predictive distribution of daily driver flows $p(\tilde{y} | \by)$ in Equation~\eqref{eq:posterior_distribution}, we are only required to input the prior $\pi(\btheta)$, the likelihood $L(\by | \btheta)$ and the recurrence relation which generates the vector of simulated driver flows $\by$ (i.e. Algorithm~\ref{alg:traffic_flow_daytypes}) into pyStan. The latter automatically calculates, for the $j$-th iteration, $j=1,\dots, J$, the vector of $N$ replicates drawn from the posterior predictive distribution:
\begin{equation}
\tilde{\pmb{y}}^{(j)} 
= \begin{bmatrix} y^{(j,1)}  \\  \vdots  \\  y^{(j,N)} \end{bmatrix}
\sim \begin{bmatrix} p(\Tilde{y}^{(j,1)} | \by)  \\  \vdots  \\  p(\Tilde{y}^{(j,N)} | \by)
\end{bmatrix}.
\label{eq:y_pred_sim}
\end{equation} 
The final output is the sequence of posterior prediction vectors $\tilde{\mathbb{Y}} = \{ \tilde{\pmb{y}}^{(1)}, \dots, \tilde{\pmb{y}}^{(J)} \}$.

\section{Bayesian Gamma regression of the passenger waiting times} \label{sec:waiting_times_model}

From the perspective of a passenger in a carpooling service, a pertinent measure of the service quality is the waiting time for a driver to arrive after the carpooling request is made. In a stochastic matching carpooling service, the daily driver flow is the predominant factor in determining this waiting time, unlike for deterministic services where it plays a minor role. So the analysis of driver flow from the previous section plays an important role in passenger waiting time prediction, as illustrated in the flowchart in Figure~\ref{fig:model_structure}. Established methods for waiting time prediction tend to be frequentist approaches based on Poisson driver arrivals, see \cite{ray2014planning,papoutsis2020door} which have encountered varying degrees of success. The aim of the section is to introduce more accurate Bayesian regression models which rely less on the Poissonian assumptions on the driver flows. 

For simplicity, we assume that a passenger can only make one request at a time for themselves only at a carpooling meeting point, and the drivers embark can only one passenger in their vehicle in the order that the passenger requests are made. 

\begin{figure}[!ht]
\centering
\includegraphics[width=\textwidth]{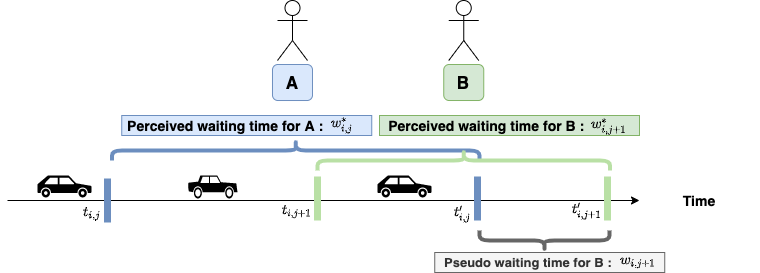} 
\caption{Perceived and pseudo waiting times for the case of two passengers at a carpooling meeting point. Passenger A is at the head of the queue so their perceived waiting time (blue brace) coincides with their pseudo waiting time. For passenger B their pseudo waiting time (grey brace) is the difference between their departure and  the departure of the previous passenger A, which is shorter than their perceived waiting time (green brace).}
\label{fig:pseudo-waiting-times}
\end{figure}

For day $i$, let $y_i$ be the daily traffic flow, and $w_{i,1}, \dots, w_{i,n_i}$ be the waiting times for the $n_i$ passengers who make a carpooling request at time $t_{i,1} < \dots < t_{i,n_i}$ respectively. Let  $t'_{i,j}$ be the driver arrival times for the passenger request at time $t_{i,j}, i=1, \dots, N$ and $j=1, \dots, n_i$.  The perceived waiting time for passenger request at time $t_{i,j}$ is  $$  w^*_{i,j} = t^{'}_{i,j} - t_{i,j}$$ and the \textit{pseudo} waiting time is $$ w_{i,j} = t^{'}_{i,j} - \max (t_{i,j}, t^{'}_{i,j-1}) $$ with the convention $t^{'}_{i,0} = t_{i,1}$ for the first passenger on day $i$. Figure~\ref{fig:pseudo-waiting-times} illustrates the difference between the perceived and pseudo waiting time for the case of two passengers A, B who are both not the first passenger of the day. Passenger A arrives first and is the $j$-th, with $j>1$, passenger of day $i$,  and makes a carpooling request at time $t_{i,j} $. Passenger B arrives immediately afterwards and is  $(j+1)$-th passenger with request time $t_{i,j+1}$. Suppose that there are at least two drivers en route to embark these passengers, and who have not received any passenger requests before passenger A's request. The first driver arrives at $t'_{i,j} > t_{i,j+1}$ (i.e. after passenger $B$'s request time) and the second driver at $t'_{i,j+1}$. The perceived waiting time for the passenger A is denoted $w^*_{i,j} = t'{i,j} - t_{i,j}$ (the blue brace in Figure~\ref{fig:pseudo-waiting-times}) and for the passenger B is $w^*_{i,+1j} = t'_{i,j+1} - t_{i,j+1}$ (the green brace).  The pseudo waiting for passenger A is $w_{i,j} = w^*_{i,j}$ since they are at the front of the queue, and for passenger B is $w_{i,j+1} = t'_{i,j+1} - t'_{i,j}$ is the grey brace. The pseudo waiting time for passenger B is the difference between their departure and  the departure of the previous passenger A, and this is shorter than the perceived waiting time $w^*_{i,j+1}$.

We focus on the the pseudo waiting times rather than the perceived waiting times in our model.   
From Figure~\ref{fig:pseudo-waiting-times}, we observe that perceived waiting times $w_{i,j}^*$ and  $w_{i,j+1}^*$ for Passengers A and B overlap, whereas the pseudo waiting times $w_{i,j}=w_{i,j}^*$ and $w_{i,j+1}$, by construction, do \textit{not} overlap. The overlapping nature of the interval processes that determine the perceived waiting times renders the problem of their prediction non-identifiable and that is why we have introduced the pseudo waiting times.
Moreover, it is  possible to predict the perceived waiting times from the pseudo waiting times, given  known passengers behaviours, e.g. there is already a passenger waiting for a car since $t$ minutes before the arrival of another passenger. Thus if `waiting time' is employed without any qualifier, it is assumed to be the pseudo waiting time.

\begin{figure}[!ht]
\centering
\includegraphics[width=\textwidth]{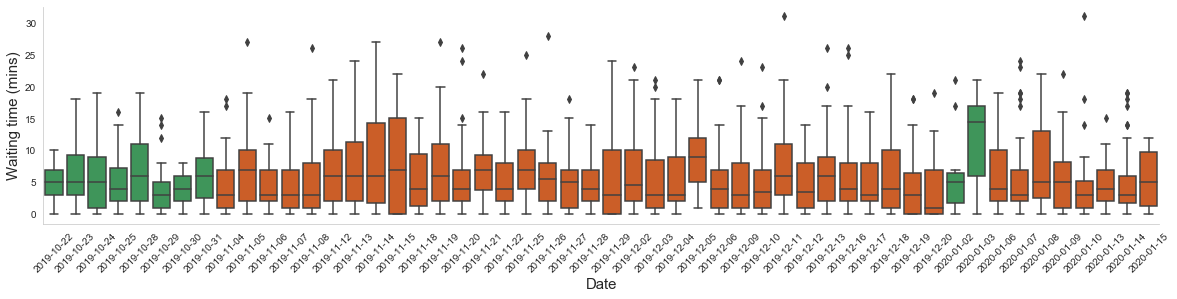} 
\caption{Observed pseudo waiting times (in minutes) in the Lane carpooling network from 2019-10-22 to 2020-01-15. The ordinary work days (ORD) are in orange, the school holidays (SCH) in green.}
\label{fig:total_waiting_times}
\end{figure}

Whilst the acquisition protocols for the driver GPS traces have been functioning well since the launch of the Lane carpooling service 2018-07-15, this was not the case for the passenger waiting times due to persistent technical operational difficulties for more than a year after the service launch. This leads to a highly challenging situation in which to deliver robust passenger waiting time predictions.  We focus on the observed passenger pseudo waiting times covering the period from 2019-07-25 to 2020-02-17.
In Figure~\ref{fig:total_waiting_times} are the 1500 observed passenger pseudo waiting times in the Lane carpooling service, from 2019-10-22 to 2020-01-15 (we plot a sub-sample of the total period for a better visualisation). This range of dates is different from those for the driver GPS traces since, due to operational technical difficulties from the service launch on 2018-05-15 until 2019-10-21, the  passenger waiting times were not reliably recorded so they are excluded from the analysis. The operation of the Lane service is guaranteed only for work days (including some school holidays), though this does not prevent passengers and drivers from using the service for other days, there are nonetheless far fewer carpooling requests for school holiday weekdays and there are none for the public holidays/weekends.

\subsection{Gamma regression model specification} \label{sec:waiting_times}
In the previous section, we implemented estimations of the driver flow at the daily level. For the waiting times, we wish to formulate predictions at sub-daily resolution. Let the 24 hour period of a day be divided into $S$ equal intervals $I_1 <  \dots < I_S$.  The fraction of the daily driver flow on each interval $I_s, s=1, \dots, S$ is $y_i \beta_s$, where $\beta_s \geq 0$ and $\sum_{s=1}^{S} \beta_s = 1$.
Conditional on the traffic flow $y_i$ and that passenger request times $t_{i,j} \in I_s$, we suppose that the pseudo waiting times $w_{i,j}$ are independent Gamma random variables with parameters $\nu$ and $\beta_s y_i$:  
\begin{equation}
w_{i,j} | (y_i, \bbeta, t_{i,j} \in I_s) \sim \Gamma(\nu,\beta_s y_i ) \quad \textrm{ for } i = 1\dots N \textrm{ and } j= 1 ,\dots, n_i.
\label{eq:pseudo_waiting_time_gamma}
\end{equation}
This model specification ensures that the conditional mean pseudo waiting time is 
$$
\mathbb{E}[w_{i,j}|(y_i, \bbeta, t_{i,j} \in I_s)] = \frac{\nu}{\beta_s y_i}
$$
which is consistent with our intuition of the inverse relationship between the driver flow and the waiting time. Since $\bbeta$ is constant for all $i$, then the model assumes that the relative proportions of the traffic flow in the intervals $I_1, \dots, I_S$ remain unchanged for all driver flow values. 

\begin{figure}[!ht]
\centering
\includegraphics[width=\textwidth]{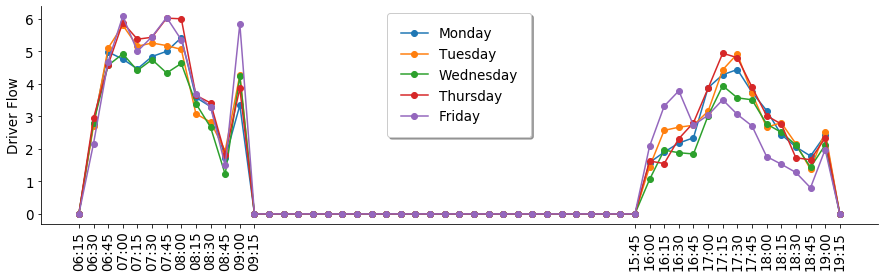}
\caption{Mean driver flows for 15 minute interval for each weekday for the Lane carpooling service, from 2018-05-15 to 2019-05-31. Monday is in blue, Tuesday in orange, Wednesday in green, Thursday in pink, Friday in violet.}
\label{fig:lane_line_quarter}
\end{figure}

In Figure~\ref{fig:lane_line_quarter} are the mean observed daily traffic flows for each weekday from the Lane carpooling service, where the day is divided into 15 minute intervals ($S=96$). Since the service operating hours are 06:00 -- 09:00 and 16:00 -- 19:00, there are few drivers outside them.  Each dot in the figure is the mean number of drivers for each 15 minute interval for each week day from 2018-05-15 to 2019-05-31. Each week day has a similar shape so this gives some empirical justification for supposing a constant $\bbeta$ for all days. 

A  Dirichlet distribution is a natural choice as a prior distribution on the coefficients $\bbeta$: $\bbeta \sim \mathrm{Dir}(S, \balpha)$ where $\balpha = (\alpha_1, \dots, \alpha_S)$ are the concentration parameters, since it imposes the constraint $\sum_{s=1}^{S} \beta_s = 1$ on the coefficients.  

The corresponding Dirichlet density is
\begin{align*}
    p (\bbeta) = \frac {1}{\mathrm{B}(\balpha)} \prod\limits_{s=1}^S \beta_s^{\alpha_s-1}
\end{align*}
where $B(\balpha) = \prod^S_{s=1} \Gamma(\alpha_s) \big/ \Gamma \big(\sum^S_{s=1} \alpha_s\big)$ and $\Gamma(x) = \int _0^{\infty} u^{x-1}e^{{-u}} du$. 
Nonetheless, for the situations where we cannot assure that the sum of the $\bbeta$ coefficients is always 1, then a non-informative prior (i.e. the Lebesgue measure on $\mathbb{R}_+^S$) is preferred. 
The GPS driver traces collected by Ecov from the mobile application represent an incomplete subset of the complete driver population of interest as they exclude (a) the drivers who are registered in the Lane carpooling service, but do not share their geolocation with Ecov, and (b) the drivers who are currently not registered but are potential participants in the carpooling service. So in this case of an incomplete driver population, the $\bbeta$ coefficients do not necessarily sum to 1.

The $\bbeta$ vector allows us to rebuild the temporal distribution of the traffic flow within a day from an aggregated daily driver flow. In the cases when the driver flow can be observed at a sub-daily level, we still prefer to apply our multi-level moving average model in Equation~\eqref{eq:traffic_model} to predict a daily driver flow as (i) it improves the robustness  and (ii) it is straightforward to change the temporal resolution $I_1, \dots, I_S$ of the waiting time predictions without having to re-generate the driver flows.

Let $\bt_i = (t_{i,1}, \dots, t_{i,n_i})$ be the vector of the $n_i$ observed passenger carpooling request times for the day $i \in \{1,\dots,N\}$, $\bt = (\bt_1,\dots,\bt_N)$ be all observed passenger carpooling request times, and likewise for the passenger pseudo waiting times $\bw_i$ for day $i$, $\bw$ for all days. Also, let the vector $\by = (y_1,\dots,y_N)$ be the observed driver flows for all days. It is reasonable to assume that the waiting times are mutually independent conditionally to $(\bbeta , \by , \bt)$. The conditional likelihood of the passenger waiting times is thus given by the joint density of $\bw$ given $(\bbeta , \by , \bt)$
$$L(\bow | \bbeta, \by, \bt) =  \prod_{i=1}^N p(\bow_i | \by ,\bt ,\bbeta)  =\prod_{i=1}^N p(\bow_i | y_i ,\bt_i ,\bbeta)   $$
since the conditional density of $\bow_i$ given the driver flow $y_i$, passenger carpooling request times $\bt_i$ and the coefficient $\bbeta$ is 
\begin{align*} 
    p(\bw_i | \bbeta, y_i , \bt_i) = \prod_{s=1}^S \prod_{\{j:t_{i,j} \in I_s\}}  \frac{(\beta_s y_i)^{\nu}}{\Gamma(\nu)}w_{i,j}^{\nu-1} \exp(-\beta_s y_i w_{i,j}).
\end{align*}
Then we obtain the posterior  density of $\bbeta$, using a non-informative prior on $\bbeta$, as 
$$\pi(\bbeta | \by, \bt, \bw )  
     \propto   \prod_{i=1}^N  \prod_{s=1}^S \prod_{\{j:t_{i,j} \in I_s\}}  \frac{(\beta_s y_i)^{\nu}}{\Gamma(\nu)}w_{i,j}^{\nu-1} \exp(-\beta_s y_i w_{i,j}) \mathbf{1} \{ \bbeta \in \mathbb{R}^S_+\}. 
    $$

Let $\tilde{w}_s $ be the pseudo waiting time for a future, unobserved day for a passenger who makes a carpooling request in the  time interval $I_s$. 
If we observe a new daily driver flow $\tilde{y}$, then the posterior predictive distribution of the waiting time $\tilde{w}_s $ is  
\begin{align} \label{eq:predictive_distribution_final}
    p(\tilde{w}_s | \tilde{y},\by,\bw) & =
    \int_0^1
    p(\tilde{w}_s| \tilde{y} ,\beta_s )  \pi(\beta_s | \by , \bw , \bt )  \, d\beta_s .
\end{align}
If $\tilde{y}$ is not observed, then 
the predictive distribution becomes 
\begin{align} \label{eq:predictive_distribution_final2}
    p(\tilde{w}_s | \by,\bw) & =
    \int_0^\infty \left[ \int_0^1
    p(\tilde{w}_s| \tilde{y} ,\beta_s )  \pi(\beta_s | \by , \bw , \bt)  \, d\beta_s \right] \, p(\tilde y|\by)  \, d\tilde y 
\end{align}
where $p(\tilde y|\by)$ is defined in Equation~\eqref{eq:pred_distri_y}. 
Equation~\eqref{eq:predictive_distribution_final} applies when we wish to make a prediction for the current day where we have observed a driver flow, and Equation~\eqref{eq:predictive_distribution_final2} for a future day where we have not yet observed the driver flow. 
Finally we wish to predict the waiting time for all the time intervals $I_1, \dots, I_S$ so we collate them into an $S$-vector

  $\left( p(\Tilde{w}_1 | \tilde y,\by,\bw), \dots, p(\Tilde{w}_S | \tilde y,\by,\bw) \right)$,
or analogously with $p(\tilde{w}_s | \by,\bw), s=1, \dots, S$. 

\subsection{Simulation algorithms for passenger waiting times}

Algorithm~\ref{alg:pseudo_wt_dist} simulates the passenger pseudo waiting times in Equation~\eqref{eq:pseudo_waiting_time_gamma} for a sequence of days. The inputs are the number of days $N$, the day types coefficients $\btheta$, the autoregression order $K$, the error variance $\sigma^2_\varepsilon$, the first shape parameter for the Gamma distribution $\nu$, the $S$ regression parameters $\bbeta$, and the number of replicates of waiting times $J$.  The output are $J$ replicates of a pseudo waiting times for each time interval $I_s, s=1,\dots,S$, for each day $i=1, \dots, N$. The \textsc{TrafficFlowDT} procedure (Algorithm~\ref{alg:traffic_flow_daytypes}) is called outside of the replicates loop so that each day has one driver flow, and all waiting times are simulated from this same daily driver flow.

\begin{algorithm}[!ht]
\textbf{procedure} \textsc{WaitingTime}($N, \btheta, K,\sigma^2_\varepsilon, \nu, \bbeta, J$) \\
$S \longleftarrow \textsc{Len}(\bbeta)$ \\
\For{i $\mathbf{in}$ 1:N}{
$Y[i] \longleftarrow \textsc{TrafficFlowDT}(i, \btheta, K, \sigma^2_\varepsilon)$ 
}
\For{j $\mathbf{in}$ 1:J}{
 \For{i $\mathbf{in}$ 1:N}{
     \For{s $\mathbf{in}$ 1:S}{
        $\pmb W^{(j)}[i,s] \longleftarrow \Gamma(\nu,\beta_s Y[i])$
    }
  }
}
\textbf{return}: $\bW^{(1)}, \dots , \bW^{(J)}$ sequence of waiting time matrices

\caption{Passenger pseudo waiting times}
\label{alg:pseudo_wt_dist}
\end{algorithm}

An iteration of the nested loop in Algorithm~\ref{alg:pseudo_wt_dist} results in a single $N \times S$ matrix of pseudo waiting times  drawn from the appropriate Gamma distributions:
$$
\bW^{(j)}
    \sim \begin{bmatrix}
           \Gamma (\nu,\beta_1 y_1)  & \hdots  & \Gamma (\nu,\beta_S y_1) \\
           \vdots  & & \vdots \\
           \Gamma (\nu,\beta_1 y_N)  & \hdots  & \Gamma (\nu,\beta_S y_N)
    \end{bmatrix}.
$$
These are then iterated $J$ times and collated into the sequence to be the output from Algorithm~\ref{alg:pseudo_wt_dist}
$$
\mathbb{W} = 
\{\bW^{(1)}, \dots , \bW^{(J)}\}
    = \left\{ 
    \begin{bmatrix}
           w^{(1)}_{1,1} & \hdots & w^{(1)}_{1,S} \\
           \vdots  & & \vdots \\
           w^{(1)}_{N,1} & \hdots &  w^{(1)}_{N,S}
    \end{bmatrix}
     , \dots ,
     \begin{bmatrix}
           w^{(J)}_{1,1}  & \hdots &  w^{(J)}_{1,S} \\
           \vdots  & & \vdots \\
           w^{(J)}_{N,1} & \hdots &  w^{(J)}_{N,S}
    \end{bmatrix}\right\}.
$$

As Equation~\eqref{eq:predictive_distribution_final} generates only a single posterior prediction $\tilde{w}_s$ for a time interval $I_s$, we collate these $\tilde{w}_s$ for $s=1, \dots S$ into an $S$-vector, and in turn collate $N$ of these $S$-vectors of posterior prediction distributions row-wise into a $N\times S$ matrix.  
To carry out the complicated integration and then a random draw from the posterior predictive distribution of $p(\Tilde{w}_s | \tilde y,\by,\bw)$ in Equation~\eqref{eq:predictive_distribution_final}, we are only required to input the posterior predicted value of the driver flow $\tilde{y}$ (Equation~\eqref{eq:pred_distri_y}), the recurrence relation which generates the vector of simulated driver flows $\by$ (Algorithm~\ref{alg:traffic_flow_daytypes}), and  the recurrence relation which generates the vector of simulated passenger pseudo waiting times $\bw$ (Algorithm~\ref{alg:pseudo_wt_dist}) into pyStan. 
The latter automatically simulates from this  $N \times S$ matrix distribution:
$$
    \tilde{\bW}^{(j)} 
    = \begin{bmatrix}
     \tilde{w}^{(j)}_{1,1}  & \dots &  \tilde{w}^{(j)}_{1,S}\\
     \vdots & & \vdots \\
     \tilde{w}^{(j)}_{1,N}  & \dots &  \tilde{w}^{(j)}_{N,S}
    \end{bmatrix}
    \sim 
    \begin{bmatrix}
     p(\Tilde{w}^{(j,1)}_1 | \tilde y,\by,\bw) & \dots & p(\Tilde{w}^{(j,1)}_S | \tilde y,\by,\bw) \\
     \vdots  & & \vdots \\
     p(\Tilde{w}^{(j,N)}_1 | \tilde y,\by,\bw) & \dots & p(\Tilde{w}^{(j,N)}_S | \tilde y,\by,\bw)
    \end{bmatrix}
$$
for the $j$-th iteration, $j=1,\dots, J$. The sequence of the matrices of replicated posterior predictions is $\mathbb{\tilde W} = \{ \pmb{\tilde W}^{(1)}, \dots, \pmb{\tilde W}^{(J)}) \}$.

\section{Model validation}
\label{sec:validation}

\subsection{Simulated passenger pseudo waiting times}
\label{sec:sims}

Since the Lane carpooling data set dates from 2018, for the simulations, we set the initial day $i=1$ to be 2018-01-01, and the work days (ORD), school (SCH) and public holidays/weekends (PWE) to be those observed in Lyon, France. For the simulation algorithms, the parameters are: the number of days is $N=365$, the day types coefficients is $\btheta = (0.333,0.33,0.331,1,1,1)$, the autoregression order is $K=3$, the error variance is $\sigma^2_\varepsilon = 5$, the 24 hour period is divided in $S=8$ equal intervals of 3 hours, the first Gamma shape parameter is $\nu=7$, the Gamma regression parameters are $\bbeta =$ (0.012, 0.01, 0.011, 0.013, 0.018, 0.016, 0.017, 0.019), and the number of replicates is $J=10$ which corresponds to the number of observed waiting times per time interval. These parameter values produce simulated data which is comparable to those observed in the Lane carpooling service.

\begin{figure}[!htp]
\centering
\includegraphics[width=0.7\textwidth]{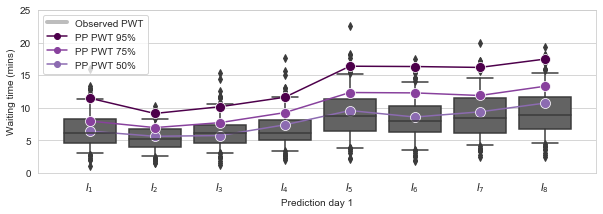}
\includegraphics[width=0.7\textwidth]{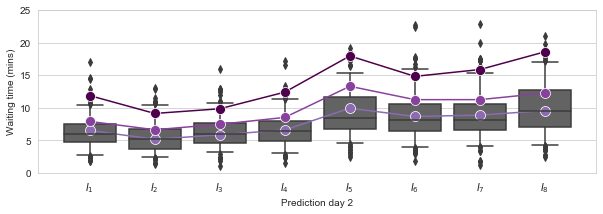}
\includegraphics[width=0.7\textwidth]{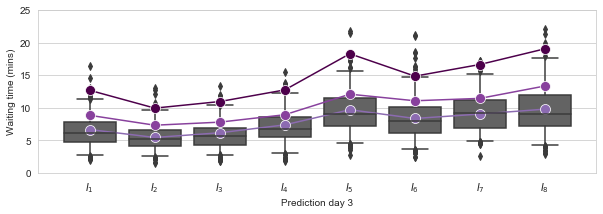}
\includegraphics[width=0.7\textwidth]{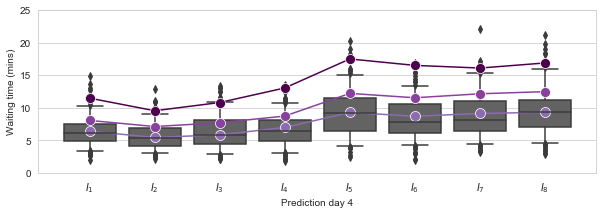}
\includegraphics[width=0.7\textwidth]{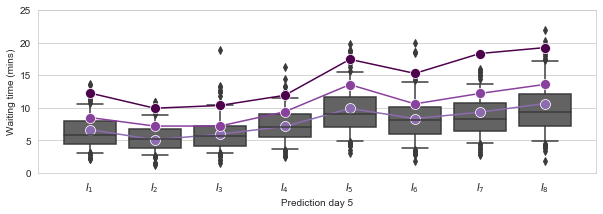}
\caption{Predictions of pseudo waiting times for 3-hourly intervals, for all $\tilde{N}$ prediction days. The observed waiting times (Observed PWT) are the grey box plots, and the 50\%, 75\%, 95\% quantiles of the posterior predicted waiting times (PP PWT) are the light, medium and dark purple circles.}
\label{fig:post_emp_quantiles}
\end{figure}

We generate one simulated data set of $N=365$ days, each with one daily driver flow $y_i,i=1,\dots,N$ (Algorithm~\ref{alg:traffic_flow_daytypes}), and $J=10$ passenger pseudo waiting time $N\times S$ matrices $\mathbb{W} = \{ \bW^{(1)}, \dots, \bW^{(J)}\}$ (Algorithm~\ref{alg:pseudo_wt_dist}), and the corresponding $N \times S$ posterior prediction matrices to form $\tilde{\mathbb{W}} = \{\tilde{\bW}^{(1)}, \dots, \tilde{\bW}^{(J)}\}$. The data from these $N=365$ days from 2018-01-01 to 2018-12-31 form the reference training data set. With the same parameters, we simulate a further $\tilde{N}=5$ days (2019-01-01 to 2019-01-05) of the data, which forms the oracle test data set of the $\tilde{N} \times S$ matrices to form $\mathbb{W}_{\test} = \{ \bW^{(1)}_{\test}, \dots, \bW^{(J)}_{\test}\}$. Furthermore, from the training data only (i.e. we do not take into account the $\mathbb{W}_{\test}$), for these same extra $\tilde{N}$ days, we generate the corresponding $\tilde{N} \times S$ posterior prediction matrices to form $\tilde{\mathbb{W}}_{\test} = \{\tilde{\bW}^{(1)}_{\test}, \dots, \tilde{\bW}^{(J)}_{\test}\}$. For brevity we have omitted the equivalent comparison of the driver flows and focus on the passenger waiting times for these simulated data: we make a more thorough comparison of both driver flows and passenger waiting times for the empirical data in the sequel.  
Whilst the daily traffic flows can be summarised by a single scalar, for the passenger waiting times, we focus on the temporal profiles, over the $S=8$ periods of a day, of the waiting times.
In Figure~\ref{fig:post_emp_quantiles} are the quantiles of the waiting times for all time intervals $I_1, \dots, I_S$, for all days $i=1, \dots, \tilde{N}$ in the test phase. The grey box plots are of 
$\mathbb{W}_{\test, i,s}$ and the light, medium and dark purple circles superimposed over the box plots are the 50\%, 75\%, 95\% quantiles of $\tilde{\mathbb{W}}_{\test,i,s}$. Recall that for operational purposes of the Lane carpooling service, short term prediction for the coming week is sufficient. This is verified by the close of the  quantiles of the posterior predicted pseudo waiting times with their observed values for all $\tilde{N}=5$ prediction days.

\begin{figure}[!ht]
\centering
\includegraphics[width=0.8\textwidth]{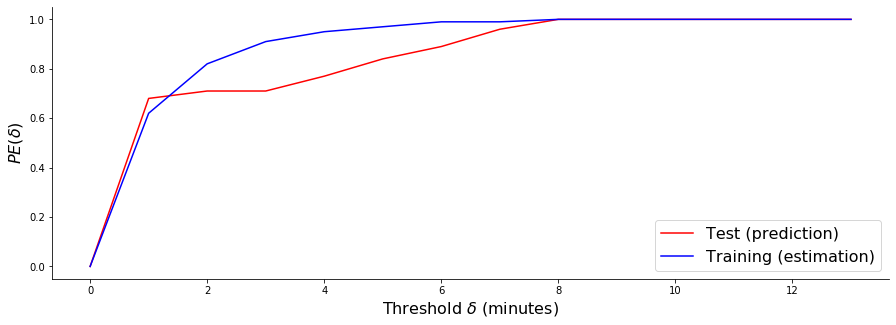}
\caption{Evolution of the PE metric of the observed and posterior predicted daily driver flows, as a function of the threshold $\delta$. The blue curve is for the training phase, and the red curve for the test phase.}
\label{fig:pe-simu}
\end{figure}

From a passenger point of view, whilst the magnitude of waiting time is important as a perception of the service quality, it is equally important that these posterior predicted waiting times be as close to the observed ones, whatever their magnitude. For example, suppose that a driver arrives after 12 minutes a passenger makes a carpooling request. In this case, a prediction of 15 minutes is better than 5 minutes since the former is closer to, but longer than, the observed waiting time than the latter. Therefore we propose the following metric to measure these discrepancies for a given threshold $\delta$:  
\begin{align}
\label{metric:custom}
    \PE(p(\mathbb{W}_{\test}), p(\tilde{\mathbb{W}}_{\test}); \delta) = \frac{1}{J\tilde{N}S} \sum \limits_{i=1}^{\tilde{N}} \sum \limits_{s=1}^S \sum \limits_{j=1}^{J}  \mathbf{1}\{|\Bar{\tilde{w}}_{\test,i,s} - w_{\test,i,s}^{(j)} | < \delta\} 
\end{align}
where $\Bar{\tilde{w}}_{\test,i,s} = \tfrac{1}{J} \sum_{j=1}^{J} \tilde{w}_{\test,s}^{(j,i)}$ is the mean of the posterior predicted waiting times distribution for day $i, i=1,\dots,\tilde{N}$ and time interval $I_s, s=1,\dots, S$. This metric, as a function of $\delta$,  illustrated in Figure~\ref{fig:pe-simu}, during both the training phase $\PE(p(\mathbb{W}), p(\tilde{\mathbb{W}}); \delta)$ (blue curve) and the test phase $\PE(p(\mathbb{W}_\test), p(\tilde{\mathbb{W}}_\test); \delta)$ (red curve). The test predictions are more accurate than the training predictions for small values of $\delta < 2$ minutes since red PE curve is above the blue PE curve in this interval. This reverses for $\delta$ between 2 and 8 minutes, and after 8 minutes, both curves level off at 1. Thus the posterior predictions from our proposed Bayesian hierarchical model can have robust prediction performance. 

\subsection{Empirical data from the Lane carpooling service}
\label{sec:application}
Our objective is the employ the two-stage Bayesian hierarchical model (Algorithms~\ref{alg:traffic_flow_daytypes} and \ref{alg:pseudo_wt_dist}) to predict a passenger pseudo waiting time (distribution) for hourly intervals $I_s, s=1, \dots, S$, with $S=24$. For the Lane carpooling service, it is sufficient to provide the upcoming week's predictions at the beginning of the week. The predicted daily driver flows from the first hierarchical model are input into the second hierarchical model to produce predicted passenger pseudo waiting times. The latter are then compared to the observed pseudo waiting times of the passenger carpooling requests from the same period.

\subsubsection{Daily driver flows}

We have the GPS traces (approximately 5 000 traces) for the 382 days from 2018-05-15 (service launch) to 2019-05-31 (beginning of following summer holiday period), which we divide into different training and test data sets of varying sizes depending the objectives of the analysis. 
We first apply the preprocessing, as outlined in \citep{papoutsis2020door}, to the driver GPS traces to the convert into data format suitable for computing the daily driver flows $y_{\train,i}, i=1, \dots, N$ for the training and $y_{\test, i}, i=1, \dots, \tilde{N}$ for the test phases. We vary $N$ whilst maintaining $\tilde{N}=7$ to test various scenarios in different periods of the year. 
To investigate the prediction accuracy of our proposed models, we divide our complete data set of into 6 different pairs of training phases with varying $N$ (starting from 2018-05-15) and test phases with $\tilde{N}=7$ always. In each case we select a test week with certain characteristics as outlined in Table~\ref{tab:test_set_scenario}. The first column are the dates (inclusive) of the test week, the second column are the day types in the test week, the third column are are the dates of the training weeks starting from 2018-05-15 to the previous day of the test week, and the fourth column is the number of training days ($N$).

\begin{table}[!htp]
\centering
\caption{Training-test scenarios for daily driver flows. The first column are the dates of the test week ($\tilde{N}=7$), the second is the day types in the test week, the third are the dates of the training weeks and the fourth column is the number of training days $N$.}
\label{tab:test_set_scenario}
\resizebox{\textwidth}{!}{%
\begin{tabular}{lcccc}
& Test week & Test week day types & Training weeks & \#training days ($N$) \\ \hline
\#1 & 2019-01-14 – 2019-01-20 & \begin{tabular}[c]{@{}c@{}}All ORD after holiday\\ period (PWE/SCH)\end{tabular}             & 2018-05-15 – 2019-01-13 & 244                   \\
\#2 & 2019-02-25 – 2019-03-03 & All SCH                                                                                      & 2018-05-15 – 2019-02-24 & 286                   \\
\#3 & 2019-04-29 – 2019-05-05 & \begin{tabular}[c]{@{}c@{}}All ORD except 1 PWE\\ (2019-05-01)\end{tabular}                  & 2018-05-15 – 2019-04-28 & 349                   \\
\#4 & 2019-05-06 – 2019-05-12 & \begin{tabular}[c]{@{}c@{}}All ORD except 1 PWE\\ (2019-05-08)\end{tabular}                  & 2018-05-15 – 2019-05-05 & 356                   \\
\#5 & 2019-05-13 – 2019-05-19 & \begin{tabular}[c]{@{}c@{}}All ORD except 1 PWE\\ (transport strike 2019-05-16)\end{tabular} & 2018-05-15 – 2019-05-12 & 363                   \\
\#6 & 2019-05-20 – 2019-05-26 & All ORD                                                                                      & 2018-05-15 – 2019-05-19 & 370                  
\end{tabular}%
}
\end{table}

In addition to our proposed Bayesian hierarchical multi-level (BHML) predictions, we compute predictions from two other models: baseline frequentist (BASE) and the Bayesian Prophet model (PROP). The baseline frequentist model has multi-levels like our BHML, but without the Bayesian  moving average structure. To account for the for public/school holidays, as proposed by \cite{gould2008forecasting}, if day $i$ is not a school/public holiday then the average is calculated over all previous days with the same day of week as day $i$; and if day $i$ is a public holiday, then the average is over all previous public holidays. That is,  
\begin{equation}
y_i =  \frac{1}{|T_d(i)|} \sum \limits_{k \in T_d(i)} y_{i-k} \mathbf{1} \{\DT'(i)\neq\HOL\} + \frac{1}{|T_\HOL(i)|} \sum\limits_{k \in T_\HOL(i)} y_{i-k} \mathbf{1} \{ \DT'(i)=\HOL\}  + \varepsilon_i
\label{eq:baseline_estimator}
\end{equation}
where we collapse the day types function DT to  
$\DT'(i) = \mathrm{ORW}$ if $i$ is an ordinary work day or a weekend day, $\DT'(i) = \HOL$ if $i$ is a school or public holiday; DN is day of week number function, $\DN(i)=1$ if $i$ is a Monday, $\DN(i)=2$ if $i$ is a Tuesday etc; and $T_d(i) =  \{k: k < i, \DN(i-k) = d\}$ is the set of prior days with the same day of week as day $i$,  and  $T_\HOL(i) = \{k: k < i, \DT'(i-k) = \HOL\}$ is set of public holidays before day $i$.

The Bayesian Prophet model, devised by \cite{taylor2018forecasting,prophet}, is an additive model with three components:  
\begin{align}
y_i &= g(i) + s(i)+ h(i) + \varepsilon_{i} 
%g(i) & = (k + \ba(t)^\top \bdelta)t + (m + \ba(i)^\top\bgamma)  \nonumber\\
%s(i) & = \boldeta(i)^\top \balpha \nonumber \\
%h(i) & = \bh(i)^\top \bkappa.
\label{eq:prophet2}
\end{align}
where $g(i)$ is the trend, $s(i)$ is the seasonality, and $h(i)$ is the holidays effect. 
The linear trend is $g(i) = (k + \ba(i)^\top \bdelta)i + (m + \ba(i)^\top \bgamma)$
where $k$ is the growth rate, $m$ is the offset, $\ba$ is the change point indicator, $\bdelta$ is the growth rate adjustment, and $\bgamma$ is the piece-wise continuity adjustment to ensure that $g$ is continuous. The seasonality component is a Fourier decomposition
%, following from \cite{bruhns2005non,launay2015construction,taylor2018forecasting}, 
$s(i) = \sum_{\ell=1}^L [ \alpha_{\ell} \cos(2\pi \ell i/P) + \beta_{\ell} \sin (2\pi \ell  i/P)]$
where $(\alpha_{\ell},\beta_{\ell})$ are the Fourier coefficients,  
$L$ is the number of Fourier coefficients and $P$ is the period (in days). 
The holiday effect is $h(i) = \bh(i)^\top \bkappa$ where, say, 
$\bh(i) = (\mathbf{1}\{\DT(i)=\SCH\},  \mathbf{1}\{\DT(i)=\PWE\})$
is the vector of indicator variables of the type of holiday of day $i$, and
$\bkappa$ is the weight vector, usually equal to the all-ones vector. 
\cite{taylor2018forecasting} provide the details for the construction of the change point function $\ba(t)$ and the continuity adjustment parameter $\bgamma$. These authors set the number of Fourier coefficients to be $L=10$ for yearly cycles and $L=3$ for weekly cycles. What remains is to estimate the trend growth rate $k$, the offset $m$, the growth rate adjustments $\bdelta$ and the Fourier coefficients $\balpha$.

\begin{figure}[!ht]
\centering
\includegraphics[width=1\textwidth]{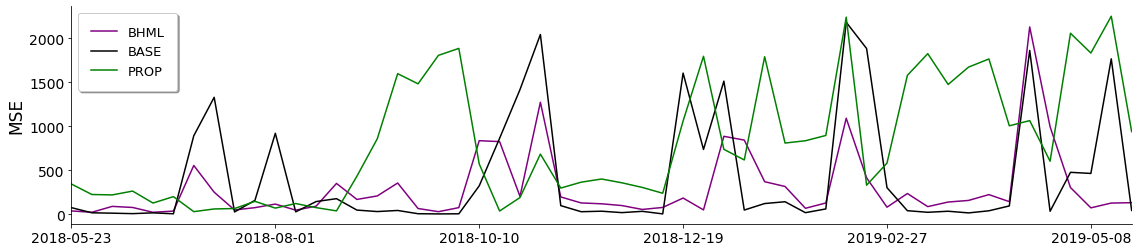}   
\caption{Evolution of the goodness-of-fit of the daily driver flow estimations over the training period  (2018-05-15 to 2019-05-19, test scenario \#\,6). Goodness-of-fit is measured by the weekly aggregated estimation MSE. Bayesian hierarchical multi-level BHML is in purple, frequentist baseline BASE in black, and Bayesian Prophet PROP in green.}
\label{fig:fitting_errrors}
\end{figure}

For the training phase of dates 2018-05-15 to 2019-05-19 (Test scenario \#\,6), we input the daily driver flows into the first hierarchical model of multi-level moving averages to produce the posterior predicted daily driver flows $\tilde{y}_i$ from Bayesian hierarchical multi-level model BHML, as well the corresponding predictions/estimations from the frequentist baseline model BASE and the Bayesian Prophet model PROP. In Figure~\ref{fig:fitting_errrors} is the evolution of the goodness-of-fit of the three different models for daily driver flow estimation (leaving out the first week 2019-05-15 to 2019-05-22 which serves as the `burn-in' period). The goodness-of-fit is measured by the MSE of the estimated and the observed daily driver flows, aggregated per week. Visually the BHML tends to have the best goodness-of-fit (smallest MSE) for most weeks. The sum of these weekly MSEs are: BASE: 421.9, PROP: 816.9, BHML: 297.2, which confirms our visual impression that the BHML achieves the best overall estimation accuracy.

\begin{figure}[!ht]
\subfloat{\includegraphics[width = 0.5\textwidth]{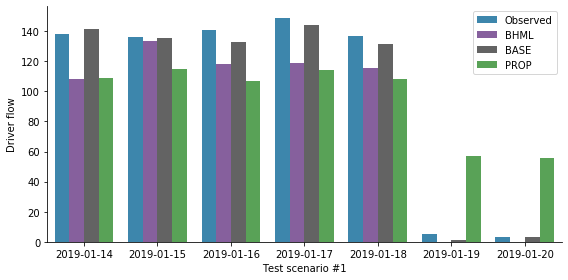}} 
\subfloat{\includegraphics[width = 0.5\textwidth]{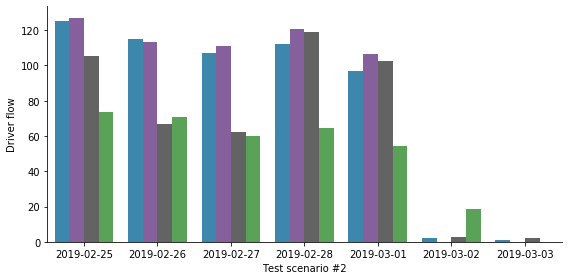}}\\
\subfloat{\includegraphics[width = 0.5\textwidth]{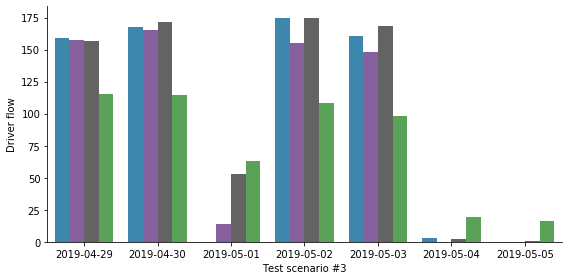}}
\subfloat{\includegraphics[width = 0.5\textwidth]{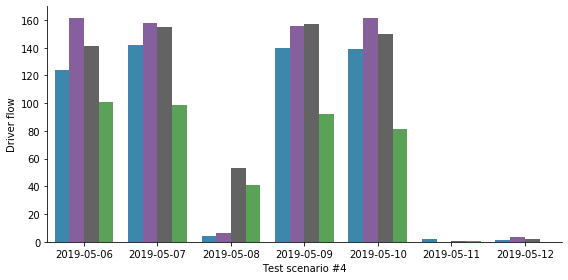}}\\
\subfloat{\includegraphics[width = 0.5\textwidth]{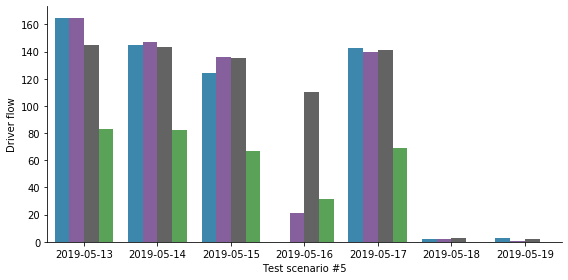}}
\subfloat{\includegraphics[width = 0.5\textwidth]{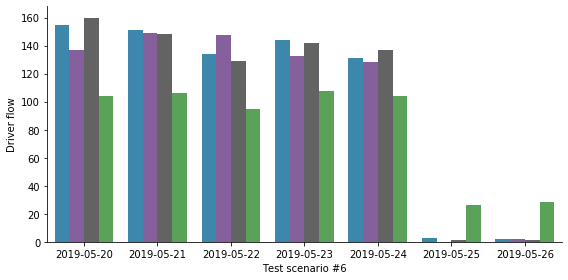}}  
\caption{Predictions of daily driver flows for the six test week scenarios. Observed daily driver flows are in blue. Bayesian hierarchical multi-level BHML are in purple, frequentist baseline BASE in black, and  Bayesian Prophet PROP in green.}
\label{fig:pred_on_test_sets}
\end{figure}

We can be confident that the Bayesian hierarchical multi-level moving average model has good estimation accuracy/goodness-of-fit, but this good performance does not necessarily translate to prediction. This non-transitivity of estimation and prediction performance is discussed in \cite{hyndmann2019}.  So for each scenario described in Table~\ref{tab:test_set_scenario}, we compute the BHML, BASE and PROP models for the training phase, and then the days of the test phase are input into each these models to yield the daily driver flow predictions. These predictions are presented in Figure~\ref{fig:pred_on_test_sets}: the Bayesian hierarchical multi-level BHML in purple, the frequentist baseline model BASE are in black, the Bayesian Prophet PROP in green; and the observed daily driver flows are in blue. The PROP predictions are mostly too low on week days and too high on weekends for all six test weeks in comparison to the observed driver flows, whilst the BASE and BHML appear to have comparable performance. 

\begin{figure}[!ht]
\centering
\includegraphics[width=0.8\textwidth]{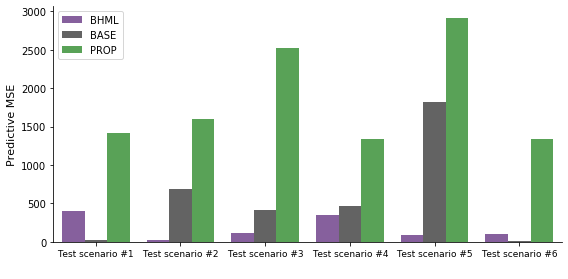}
\caption{Prediction MSE of the daily driver flow predictions for the six test week scenarios. Bayesian hierarchical multi-level BHML are in purple, frequentist baseline BASE in black, and  Bayesian Prophet PROP in green.}
\label{fig:predictive_mse_test}
\end{figure}

In Figure~\ref{fig:predictive_mse_test} are the MSEs between the observed and predicted daily driver flows: the frequentist baseline model are in black, the Bayesian Prophet PROP in green, and the Bayesian hierarchical multi-level BHML in purple. PROP is the uniformly the worst of these three models for all test weeks. BASE is the best for test scenario \#1 (all ORD after PWE/SCH period) and \#6 (all ORD) with almost zero prediction MSE, though the difference with BHML is not so large. These two test scenarios are where all days in the week are the same day type. For the other test week scenarios \#1, \#3, \#4, \#6, BHML has the smallest prediction MSE, sometimes by a large margin. For instance, the aggregated weekly MSE for the BHML model for the week starting on "2019-05-08" is 74 in contrast to the BASE and PROP model which the MSE is 464 and 1833 respectively. These test week scenarios include a day  which is a different day type to the other days within the test week, which the BHML handles the best. Overall the BHML has the best prediction accuracy for these test week scenarios.     

\subsubsection{Temporal profiles of passenger pseudo waiting times}

\begin{figure}[!htp]
\centering
\includegraphics[width=\textwidth]{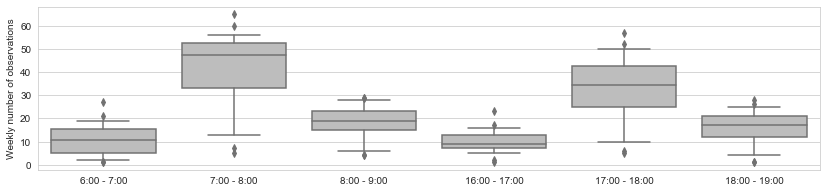}
\caption{The box plots of the weekly number of observed pseudo waiting times for each hourly interval for weekdays from 2019-07-25 to 2020-02-17.}
\label{fig:typical-week-wt-count}
\end{figure}

For the case study for simulated data in Section~\ref{sec:sims}, we could generate the oracle simulated temporal profiles to which the posterior predicted profiles could be compared. For the Lane carpooling service, since it is still in an embryonic phase of operation, there are insufficient passenger carpooling requests to robustly compute observed temporal profiles over an entire day,especially for school holidays (SCH) and public holidays/weekends (PWE) as shown in Figure~\ref{fig:total_waiting_times}. So it is only possible to form predictions for weekdays (ORD).  In Figure~\ref{fig:typical-week-wt-count} are the box plots of the weekly number of observed pseudo waiting times for each hourly interval for weekdays from 2019-07-25 to 2020-02-17, but the effective end date is 2019-05-15 since the last two days are weekend days. Although there are in total $S=24$ hourly intervals, only those 6 which correspond to the operating hours of the Lane service (06:00 -- 09:00 and 16:00 -- 19:00) contain any observed passenger waiting times.

There are a maximum of around 50-60 observed waiting times per hourly interval per week, which are not sufficient to infer robustly their distribution within each interval. To remedy this data sparsity, we aggregate a moving window of test data so for time interval $I_s$ day on $i$, we combine its observed pseudo waiting times $w_{\test,i,s}$ with those for the same time interval from the previous 5 weeks with the same day of week $\DN(i)$ and same day type $\DT(i)$, i.e. $\{ w_{\test,i-k,s} : \DT(i-k) = \DT(i), \DN(i-k) = \DN(i), k=1, \dots, 35\}$. These days added to the test data are correspondingly removed from the training data. If we aggregate the final 5 weeks to be a single test phase, then the training-test scenario is outlined in Table~\ref{tab:test_set_scenario2}. Thus we make predictions for only the last test week (2020-02-10 -- 2020-02-17), so the number of prediction weekdays remains $\tilde{N}=5$. 

\begin{table}[!htp]
\centering
\caption{Training-test scenario for passenger waiting times. The first column are the dates of the test weeks , the second is the number of observed passenger waiting times in the test weeks, the third are the dates of the training weeks and the fourth column is the number of observed passenger waiting times in the training weeks.}
\label{tab:test_set_scenario2}
\begin{tabular}{cccc}
Test weeks              & \begin{tabular}[c]{@{}c@{}}\#observed waiting \\ times in test weeks\end{tabular} & Training weeks          & \begin{tabular}[c]{@{}c@{}}\#observed waiting \\ times in training weeks\end{tabular} \\ \hline
2020-01-13 - 2020-02-17 & 520 & 2019-07-25 - 2020-01-12 & 1289                                                                                 
\end{tabular}
\end{table}

\begin{figure}[!ht]
\centering
\includegraphics[width=0.8\textwidth]{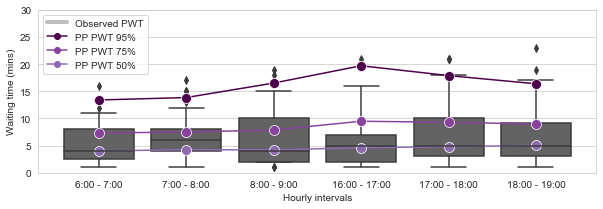}  
\caption{Predictions of passenger pseudo waiting times for hourly time intervals.  
The observed waiting times are the grey box plots, and the 50\%, 75\%, 95\% quantiles of the posterior predicted waiting times are the light, medium and dark purple circles.}
\label{fig:observed-posterior-h}
\end{figure}

In Figure~\ref{fig:observed-posterior-h} are the box plots of the observed pseudo waiting times and the  quantiles for the posterior predictions, for hourly intervals for the scenario in Table~\ref{tab:test_set_scenario2}. The observed pseudo waiting times are displayed as the grey box plots, and the 50\%, 75\%, 95\% quantiles of the posterior predicted waiting times are the light, medium and dark purple circles. The number of observations within each time interval is the number on the median line of the box plots.

\begin{figure}[!ht]
\centering
\subfloat{\includegraphics[width=0.8\textwidth]{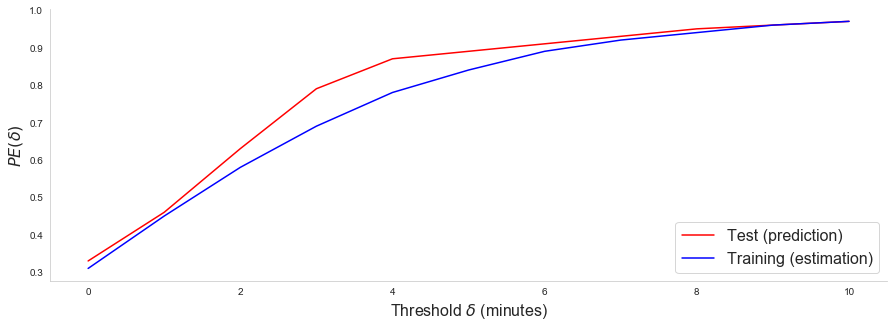}}
\caption{Evolution of the PE metric of observed and BHML posterior predicted passenger pseudo waiting times, as a function of the threshold $\delta$. The blue curve is for the training phase, and the red curve for the test phase.}
\label{fig:pe-real}
\end{figure}

Lastly we consider our custom PE metric on the BHML posterior predictions:
\begin{align}
    \PE(p(\mathbb{W}), p(\tilde{\mathbb{W}}); \delta) = \frac{1}{\tilde{N}S} \sum \limits_{i=1}^{\tilde{N}} \sum \limits_{s=1}^S \sum   \mathbf{1}\{|\Bar{\tilde{w}}_{i,s} - w_{i,s} | < \delta\} 
\end{align}
where $\Bar{\tilde{w}}_{i,s} = \tfrac{1}{J} \sum_{j=1}^{J} p(w_{s}^{(j,i)}| \tilde{y}, \by,\bw)$ is the mean of the posterior predicted waiting times distribution for day $i$ and time interval $I_s$. This metric, as a function of the threshold $\delta$,  illustrated in Figure~\ref{fig:pe-real}, during both the training phase $\PE(p(\mathbb{W}), p(\tilde{\mathbb{W}}); \delta)$ (blue curve) and the test phase $\PE(p(\mathbb{W}_\test), p(\tilde{\mathbb{W}}_\test); \delta)$ (red curve). This is in contrast to the conclusion from Section~\ref{sec:sims}, since the red curve dominates the blue curve which implies that the posterior predictions are more accurate during the test phase than in the training phase. This gives us confidence that the BHML posterior predictions are robust and are not based on over-fitting on the training data.

\section{Conclusion} \label{sec:conclusion}

The main contribution of this paper is the transformation of daily driver flows to passenger waiting times for hourly intervals using a nested two-stage Bayesian hierarchical model for an operational carpooling service. The first stage is a multi-level moving average model of the daily driver flows, whose coefficient $\btheta$ with levels depending on if the current day is a work day, a school holiday or a public holiday/weekend. The second stage is a Gamma regression whose response variables are the hourly passenger waiting times, covariates are the daily drive flows from the first stage, and  regression coefficients $\bbeta$ has as many components as the number of hourly intervals. The predicted driver flows and passenger waiting times are robust going into the future, since we demonstrate that they are not due to over-fitting of observed data from an operational carpooling service. 
We have focused on the mathematically simpler case of pseudo waiting times. The Bayesian hierarchical framework that we have employed is able to generalise to the more difficult, but more realistic case of perceived waiting times. Suppose that the pseudo waiting times for two consecutive passengers are $w_1, w_2$. Then the perceived waiting time of the second passenger is $w_2^* = w_2 + (w_1-\zeta | (w_1 > \zeta))$ with $\zeta = t_2 - t_1$. It would be intractable to deduce a closed form of the distribution of $w_2^*$. Since we are able to simulate from the conditional posterior predictive distribution of the pseudo first waiting time $w_1 - \zeta | (w_1 > \zeta)$ and from the unconditional second pseudo waiting time, then it is feasible to simulate the second perceived waiting time, assuming that these two components of $w_2^*$ are independent. Since we our primary data source are the GPS driver traces, we focused on modelling the driver arrival processes and assumed to the passenger arrivals to be non-random. In a Bayesian hierarchical framework, it is straightforward to allow the passenger arrivals to also be a random process, and to analyse the resulting pseudo and perceived passenger waiting times.

Finally we made the assumption that a driver embarks only one passenger at a time, whereas it is of intense operational interest for a carpooling provider to encourage different passengers to share a single carpooling ride, as maximising the occupancy rate in private vehicles is a key objective in the progress towards carbon-neutral societies. This passenger sharing probability is able to be analysed within the Bayesian hierarchical framework.

For future works, we consider to construct an informative prior for new network of lines with small amount of available data. Another idea is to integrate the size and number of intervals $S$ into the model, and find the optimal value with Bayesian inference. 

\section*{Acknowledgements}
The authors thank Ecov for providing the data sets of driver GPS traces and passenger waiting times. The authors also thank Constant Bridon, Safa Fennia and Madeleine Zuber from Ecov, and Gérard Biau from Sorbonne University for their feedback.

\bibliographystyle{chicago}
\bibliography{references}

\end{document}